\documentclass[%
 reprint,
 amsmath,amssymb,
pra,
]{revtex4-2}

\usepackage{graphicx}
\usepackage{dcolumn}
\usepackage{multirow}
\usepackage{amsmath}
\usepackage{amssymb}
\usepackage{euscript}
\usepackage{bm}

\usepackage{xcolor}
\usepackage{lipsum}
\usepackage{hyperref}
\usepackage{qcircuit}
\usepackage{graphicx}
\graphicspath{{figures/}}
\usepackage{dcolumn}
\usepackage{bm}
\usepackage{bbold}
\usepackage[utf8]{inputenc}
\usepackage[T1]{fontenc}
\usepackage{mathptmx}
\usepackage{mathtools}
\usepackage{xcolor}
\usepackage{tikz}
\usepackage{braket}
\usepackage{wrapfig}

\definecolor{ao(english)}{rgb}{0.0, 0.5, 0.0}

\usepackage[english]{babel}

\begin{document}

\preprint{AIP/123-QED}

\title {QREChem: Quantum Resource Estimation Software for Chemistry Applications}

\author{Matthew Otten}
\email{Corresponding author email here}
\affiliation{%
 Materials and Microsystems Laboratory, HRL Laboratories, Malibu, CA 90265, USA
}

\author{Byeol Kang}
\affiliation{%
 School of Materials Science and Engineering, Gwangju Institute of Science and Technology, Gwangju, Republic of Korea
}

\author{Dmitry Fedorov}
\author{Anouar Benali}
\author{Salman Habib}
\author{Yuri Alexeev}
\affiliation{%
 Computational Science Division, Argonne National Laboratory, Lemont, IL 60439, USA
}

\author{Stephen K. Gray}
\affiliation{%
 Center for Nanoscale Materials, Argonne National Laboratory, Lemont, IL 60439, USA
}


\begin{abstract}

As quantum hardware continues to improve, more and more application scientists have entered the field of quantum computing. However, even with the rapid improvements in the last few years, quantum devices, especially for quantum chemistry applications, still struggle to perform calculations
that classical computers could not calculate. In lieu of being able to perform specific calculations, it is important have a systematic way of estimating the resources necessary to tackle specific problems. Standard arguments about computational complexity provide hope that
quantum computers will be useful for problems in quantum chemistry but obscure the true impact of many algorithmic overheads.
These overheads will ultimately determine the precise point when quantum computers will perform better than classical computers. 
We have developed QREChem to provide
logical resource estimates for ground state energy estimation in quantum chemistry through a Trotter-based quantum phase estimation approach. QREChem provides resource estimates which 
include the specific overheads inherent to problems 
in quantum chemistry by including heuristic estimates of the number of Trotter steps and number of necessary ancilla, allowing for more accurate estimates of the
total number of gates. We utilize QREChem to
provide logical resource estimates for a variety of small molecules in various basis sets, obtaining estimates in the range of $10^7-10^{15}$ for total number of T gates.
We also determine estimates for the FeMoco molecule and compare all estimates to other
resource estimation tools.


\end{abstract}
\maketitle

\section{Introduction}
Quantum chemistry is often quoted as a potential ``killer app'' for quantum computers, with grand targets such as solving nitrogen fixation~\cite{reiher2017elucidating}. While there 
is much promise for quantum computers in quantum chemistry, due to a potential for an exponential speed up in eigenvalue estimation in quantum chemistry via the quantum phase
estimation (QPE) algorithm~\cite{abrams_quantum_1999, kitaev_quantum_1997}, realistic resource estimates,
both at the logical and physical level, point to extremely large numbers of quantum gates and qubits necessary for even small systems~\cite{reiher2017elucidating,kim2022fault}. Alternate algorithms,
more suited to near-term, noisy intermediate scale quantum (NISQ)~\cite{preskill2018quantum} devices, such as the variational quantum eigensolver (VQE)~\cite{Peruzzo2014}, provide a potential reduction
in gate depth, but add additional complexity in optimization~\cite{Menickelly2023latency} and still require substantial gate depth for more interesting, classically intractable
systems. 
With fault-tolerant, error-corrected quantum computers capable of the required gate depth and numbers of qubits still potentially years away, accurate resource estimates will 
play a key role in understanding the progress of quantum algorithms and the trade-offs of various architectural choices. There already exists several tools for estimating 
resources to varying degrees of precision. For example, TFermion~\cite{casares2022tfermion} provides estimates of a wide variety of quantum algorithms for quantum chemistry
but relies on strict error bounds, sometimes greatly overestimating the resources for certain algorithms, while OpenFermion provides estimates of certain specific quantum chemistry methods and 
also provides some tools for estimating surface code overhead~\cite{mcclean2020openfermion}.  Microsoft has released a `full-stack' resource estimation
framework and tool~\cite{beverland2022assessing}, which allows for a more general resource estimation, including many potential hardware overheads.

Here, we detail QREChem, which provides accurate logical
resource estimates with a specific focus on quantum chemistry. Within QREChem we have implemented a detailed resource estimation of
the Trotter algorithm~\cite{PhysRevA.64.022319,PhysRevA.91.022311}, using heuristic, rather than worst-case, estimates for various algorithmic overheads. 
We have also included minimal implementations of error correction 
and hardware overheads. To benchmark our method, we compare our resource estimates to both TFermion~\cite{casares2022tfermion} and OpenFermion~\cite{mcclean2020openfermion} by estimating
the total number of logical T gates for various small molecules and for the larger FeMoco molecule~\cite{reiher2017elucidating}.

QREChem was developed in mind as a tool for providing realistic estimations of  resources to simulate \emph{ab initio} quantum chemistry calculations on various quantum 
computers, both existing and future ones. It serves multiple purposes, with a primary goal of eventual co-design of future quantum computers and development of 
new quantum algorithms. For example, given a molecule, QREChem will be able to predict the resources (the number of qubits, gates, fidelities, sampling rate) required 
to accurately estimate the ground state energy. Or another way around given quantum hardware and molecule, QREChem could calculate a potential success rate. The current version
of QREChem focuses on the algorithmic implementations; further development will involve adding more detailed implementations for the hardware and error correction overheads
to provide more precise estimates. 

\begin{figure}[!ht]
\begin{center}
\includegraphics[width=8cm]{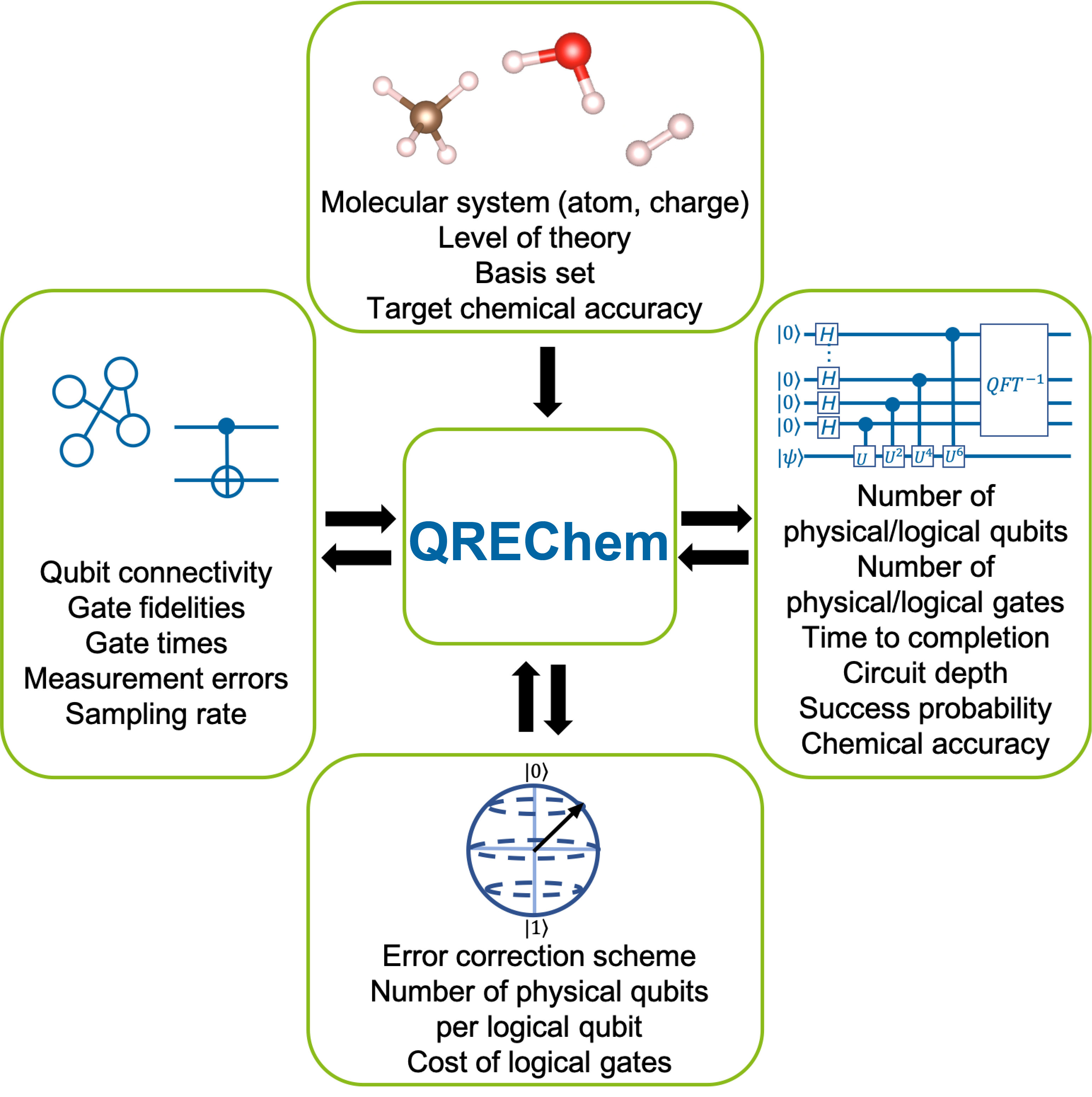}
\end{center}
\caption{Schematic diagram of QREChem.}\label{fig:1}
\end{figure}

\section{Methods}
\subsection{Design of QREChem}
QREChem is designed to allow for efficient and accurate logical resource estimates of ground state energy calculations in quantum chemistry problems. 
\autoref{fig:1} details the overall design. 
QREChem consists of several modules: the `Chemistry' module, which defines the chemical system of interest; the `Algorithm' module, which defines the quantum algorithm to be used; 
the `Hardware' module, which defines the target hardware; and the `Error Correction' module, which defines the quantum error correcting code. These modules work in tandem to produce 
the final resource estimates. Our primary
focus in this work was to provide detailed implementations of the `Chemistry' and `Algorithm' (focusing on Trotterization algorithms~\cite{PhysRevA.64.022319,PhysRevA.91.022311}) 
modules; the `Hardware' and `Error Correction' modules are relatively simple, by comparison.
In the following sections, we provide detailed descriptions of each module, as implemented in the initial version QREChem, as well as ways in which each module can be further developed.

\subsubsection{Chemistry Module}
The first step in performing quantum chemistry calculations is to generate the chemical Hamiltonian, which describes the energy operator of the molecular system in terms of the positions
of its constituent atoms. In the Chemistry module of QREChem, we generate the Hamiltonian using the self-consistent field (SCF) methods implemented in the PySCF program \cite{sun2018pyscf, sun2020recent}.

The Hamiltonian matrix elements, which represent the contributions of the various terms in the Hamiltonian, are defined in terms of one-electron integrals, $h_{pq}$, and two-electron integrals, 
$h_{pqrs}$~\cite{low2019q}. These integrals depend on the molecular properties such as the atomic coordinates, the charge, and the choice of basis set.  SCF calculations, such as 
the restricted Hartree-Fock (RHF) method, are performed to obtain the one- and two-electron integrals for the chosen molecular system. QREChem requires the definition of the chemical system of interest,
in terms of the atomic coordinates, the charge, and the target basis set.

It is important to note that the SCF calculations can be time-consuming, especially for larger numbers of atoms and larger basis sets, and can be tricky to properly converge. However, 
QREChem supports the standard \texttt{fcidump} file format, which stores the one- and two-electron integrals, allowing a user to generate these integrals using a different program and 
then interface with the other modules of QREChem. The \texttt{fcidump} file format is a widely used format for storing quantum chemistry Hamiltonians~\cite{knowles1989determinant}
and can be produced by other quantum chemistry packages, such as Gaussian~\cite{foresman1996exploring}, MolPro~\cite{werner2012molpro}, or Psi4~\cite{smith2020psi4}. 
Once the one- and 
two-electron integrals are obtained, the data is available to other modules within QREChem.

\subsubsection{Algorithm Module}
There are many proposed quantum algorithms for solving for the ground state energy in chemical problems, including quantum phase estimation (QPE)~\cite{abrams_quantum_1999, kitaev_quantum_1997}, 
the variational quantum eigensolver (VQE)~\cite{Peruzzo2014}, combinations of the two~\cite{otten2022localized,dcunha2023state}, and quantum machine learning methods~\cite{xia2018quantum}. Within each of these families of algorithms,
there are a substantial number of possible variations. In this work, we focus on QPE using Trotterization~\cite{PhysRevA.64.022319,PhysRevA.91.022311}. Here, we provide a brief overview of QPE using Trotterization.

QPE solves for the eigenvalue, $\lambda_k$, for
an eigenvector $|v_k\rangle$ of some unitary matrix, $U$. Aside from its use in ground state
energy estimation in quantum chemistry, it also finds use in Shor's prime number factoring
algorithm~\cite{shor1999polynomial} and the Hassidim--Harrow--Lloyd algorithm for matrix inversion~\cite{PhysRevLett.103.150502}. 
Given a Hamiltonian generated by the Chemistry module, the unitary matrix, $U$ is
can be written as:
\begin{equation}\label{qpe_phase}
  U |v_k\rangle = e^{i\hat{H}\tau} |v_k\rangle = e^{i 2\pi \phi} |v_k \rangle,
\end{equation}
where $\tau$ is a scale factor to map the eigenvalues of $H \tau$ onto (0, 2 $\pi$] or (-$\pi$, $\pi$]. Assuming $|v_k\rangle$ is the ground state,
the ground state energy is then mapped to the phase acquired,~$E = 2\pi \phi/\tau$, 
where units have been chosen such that $\hbar=1$. 
The computational complexity of QPE is dependent on how the unitary matrix of \autoref{qpe_phase}
is implemented. In QREChem, we focus on Trotterization~\cite{PhysRevA.64.022319,PhysRevA.91.022311},
but many other strategies, such as Taylorization~\cite{PhysRevLett.114.090502} and
qubitization~\cite{Low2019hamiltonian} have been proposed. The Trotterized version of the 
propagator, $U$, is
\begin{equation}\label{trotter}
 U = \lim_{n\rightarrow \infty} \Big(\prod_j e^{i H_j \tau/n}\Big)^n.
\end{equation}
By choosing some finite number of  Trotter steps, $n$, $U$ is only represented approximately. 
A first-order Trotter formula truncates \autoref{trotter} at some number of steps. Choosing
a sufficient number of Trotter steps for a given accuracy is important for obtaining an accurate estimate of the total resources.
Higher-order Trotter-Suzuki formulas~\cite{suzuki1993improved} can be used to decrease the number of steps at the cost of increasing
the complexity of each step.
The standard fermionic quantum chemistry Hamiltonian
has $O(N^4)$ terms. A fermion-to-spin mapping is required to implement each fermionic term on qubits. Using the Jordan--Wigner~\cite{jordan1993paulische}
transformation introduces an $O(N)$ overhead, leading to a total complexity of $O(N^5)$ for the Trotterized evolution.
The evolution of the phase is mapped to an ancilla register (introducing limits on the precision, based upon the number of ancilla)
and, using the quantum Fourier
transform (QFT)~\cite{365700}, the ground state energy can be read out. 
In realistic settings, the true eigenstate $|v_k\rangle$ is unknown and an approximation must be used. This 
introduces an additional overhead in the success probability which scales as the overlap of the approximate state, $|\phi\rangle$,
with the true eigenstate, i.e., $|\langle \phi |  v_k \rangle|^2$.

Within QREChem, we provide resource estimates for Trotterized QPE by first estimating the resources required for a single Trotter time step and then
estimating the total number of Trotter steps. The resource estimation tools within Microsoft Quantum Development Kit (QDK)~\cite{svore2018qsharp,low2019q} are 
able to efficiently provide such logical gate estimates, even for very large systems. 
We estimate the total number of Trotter steps necessary as $n_o^{3/2}$, where $n_o$ is the number of orbitals used in the Hamiltonian which is based on heuristic estimates~\cite{reiher2017elucidating,poulin2015thetrotter}.
Beyond number of Trotter steps, we also need to accurately compute the number of 
ancilla necessary to reach the a user-defined desired precision $\epsilon_p$ (which, by default, we take to be 1 mHa). This allows us to 
calculate the base number of binary digits necessary, {$n_b = -\log_2(\epsilon_p/{\Delta E_R})$, where $\Delta E_R$ is a scaling factor that estimates the spectral range (i.e., $\Delta E_R \approx E_{\mathrm{max}} - E_{\mathrm{min}}$)~\cite{dcunha2023state} and is taken to be 1 Ha (a number chosen heuristically to cover all studied molecules). The phase resulting from a QPE is given as a binary fraction and the number of bits in this 
fraction (the precision) is determined by the number of ancilla qubits used in the QPE algorithm. If the eigenvalue is not exactly representable with $n_b$ bits of precision,
the returned value will, instead, be mapped into the finite precision of $n_b$ bits, causing a chance for error. To increase the QPE success probability, additional ancilla
can be used. If the eigenvector is known precisely, the total number of ancilla, $n_a$, is a function of the desired failure probability, $p_f$,~\cite{nielsen2000quantum}
\begin{equation}
    n_a = n_b + \log_2(2 + \frac{1}{2p_f}).
\end{equation}

It is very unlikely to know the true eigenstate \textit{a priori}. More accurate formulas can be derived which take into account errors in the Trotterization, $\epsilon_t$,
as well as the true gap $\Delta E = E_1 - E_0$~\cite{li2022some}
\begin{equation}
   n_a = n_b+\log_2(2+\frac{\epsilon_t^2}{2p_f(\Delta E)^2}).
\end{equation}
Since the true gap, $\Delta E$, is generally unknown, we instead use equation-of-motion (EOM) coupled-cluster with singles and doubles (CCSD) as implemented in PySCF~\cite{sun2018pyscf}} to estimate the gap. 
In cases where CCSD becomes too expensive, other methods
with tunable cost and accuracy, such as selected configuration interaction~\cite{chien2018excited} (which can be as cheap as Hartree-Fock) can be used.
We also use a target Trotter error of chemical accuracy ($\epsilon_t = 1.6 m$Ha), rather than an observed Trotter error.

To calculate the total number of rotation gates, CNOT gates, and the total depth, we combine the estimates for a single Trotter step (using Q\# and the Microsoft QDK) with the
estimate of the total number of Trotter steps ($n_o^{3/2}$) and multiply that by the number of ancilla, $n_a$, as each ancilla will require evolution to some long time, giving,
for example, the total number of rotation gates
\begin{equation}
 n_{r,tot} = n_r n_o^{2/3} n_a,
\end{equation}
where $n_r$ is the number of rotation gates for a single Trotter step. Similar equations are used for the total depth and total number of CNOTs.

Using the same Hamiltonians generated in the Chemistry module, we utilized TFermion~\cite{casares2022tfermion} and OpenFermion~\cite{mcclean2020openfermion} to provide
comparison logical resource estimates. 
TFermion provides estimates of a variety of quantum algorithms, including variants of Trotterization~\cite{campbell2019random} and Taylorization~\cite{babbush2016exponentially}, among others. 
It uses properties of the computed Hamiltonians, such as the 1-norm, combined with analytic formulas derived from the literature.
OpenFermion provides estimation of more advanced algorithms, such as qubitization with low rank factorization~\cite{berry2019qubitization}.

Future improvements to the Algorithm module would involve the implementation of resource estimates for other evolution algorithms with explicit circuit constructions,
such as qubitization and Taylorization, as well as other algorithms, such as VQE. Furthermore, overheads relating to the preparation of sufficient overlap initial
states will also be implemented.

\subsubsection{Error Correction Module}
Quantum error correction (QEC) is an essential feature of any viable quantum computing system due to the intrinsic susceptibility of quantum systems to errors. 
These errors can be caused by a variety of factors, including decoherence and operational imperfections. 
In simple terms, QEC codes encode a logical qubit into several physical qubits, 
and through the process of measurement and classical post-processing, corrects the inevitable errors, extending the effective processing time before errors
destroy the quantum state.

There are many proposed codes within QEC. Among them, the surface code~\cite{fowler2012surface} stands out due to its high error threshold, relative ease of implementation, and 
planar geometry, which matches many proposed quantum architectures. Like any QEC code, the implementation of surface code necessitates a large number of physical qubits
to encode a single logical qubit and presents a significant overhead in terms of resource requirements. Various implementations of the surface code have a given distance, $d$, which
refers to the minimum number of physical qubits that must be affected by errors in order to cause a logical error. Within QREChem, we provide estimates of the 
QEC space and time overheads of the surface code using OpenFermion~\cite{mcclean2020openfermion}. The surface 
code cost estimator takes in the number of Toffoli gates, the number of logical qubits, a physical error rate, and the estimated surface code cycle 
time. It then estimates the total error, including error contributions from both magic state distillation (which is necessary to produce the Toffoli gates)~\cite{litinski2019magic} and due to the physical error rate. 
The physical error rate, $p_P$, logical error rate, $p_L$, and distance are approximately related via~\cite{webber2022impact,fowler2018low}
\begin{equation}\label{eq:2.17}
p_L = 0.1(100p_P)^{(d+1)/2}.
\end{equation}
The total error and space-time (number of qubits times number of seconds) is estimated for a various distances $d$ of the surface code. The best, in terms of space-time, distance $d$ estimate which has total error below a threshold $\epsilon_{sc}$ (which we by default take to be 0.1) is returned as the optimal surface code.

QREChem allows a user to input the desired total algorithmic success probability. Along with the total depth, which is estimated 
in the Algorithm module, and the physical error rate and cycle time, which is provided by the Hardware module, the error correction module provides the best surface code distance $d$. This provides an initial estimate of the QEC overhead in terms of the number of physical qubits and total runtime. 
Future module development will include more accurate estimates of the 
overhead in numbers of gates
needed for the surface code (via, for example, the methods in Ref.~\cite{litinski2019game}), as well as estimates for other QEC codes 
(such as color codes~\cite{litinski2017combining}).

\subsubsection{Hardware Module}
The underlying quantum computing architecture plays a pivotal role in the overall resource estimates. Various quantum hardware can have vastly different 
error rates, gate times, connectivities, native gate sets, etc~\cite{Menickelly2023latency}. The Hardware module captures several hardware-specific factors that can significantly affect the performance 
and resources of quantum algorithms. Within the initial release of QREChem, the Hardware module consists of a high-level description of the underlying hardware, including 
gate times and physical error rates. With this simplistic Hardware module, we estimate the logical depth by assuming that all single qubit gates can be performed in parallel batches
using $n_o$, the number of orbitals, qubits, and the CNOT gates cannot be performed in parallel ($d=\frac{n_r}{n_o} + n_c$, where $n_r$ is the number of rotation gates and $n_c$
is the number of CNOTs). To estimate the number of T gates $n_t$, which are likely to be the most expensive gate for fault-tolerant, error-corrected quantum computers~\cite{beverland2016protected,bravyi2013classification}, we
use estimates of the circuit synthesis cost of arbitrary rotations from Clifford+T gates~\cite{selinger2015efficient}
\begin{equation}
    n_t = n_r (10 + 12\log_2(\epsilon_{ss}^{-1})),
\end{equation}
where $\epsilon_{ss}$ is the synthesis error, which we take to be $10^{-9}$. We choose this value of the threshold to keep the synthesis error well below the standard $1/\sqrt{N_g}$ bound, where $N_g$ is the number of gates~\cite{hastings2016turning}, for all circuits studied. This is a tunable parameter which can be varied by a user.
Total runtime is computed by interfacing with the Error Correction module, which requires a surface code cycle time, physical error rate, number of Toffoli gates (which is related to the number of T gate) and number of logical qubits. In the initial release of QREChem, we use experimentally demonstrated cycle times to abstract away the hardware details.

Future development of this module will incorporate underlying connectivity, specific noise models, and compilation to native gate sets, along with gate times, to computer the surface code cycle time, as it is evident from the results 
of this work that a fault-tolerant quantum computer will be required to execute QPE quantum circuits.

\section{Results}

\subsection{Benchmark Molecules} 
Using QREChem, TFermion, and OpenFermion, we estimated the required logical quantum resources for the Trotter-based QPE algorithm to compute the ground state properties of various small molecules, 
including H$_{2}$, HF, H$_{2}$O, NH$_{3}$, CH$_{4}$, Be$_{2}$, and C$_{2}$. The geometries of these molecules were taken from the NIST database \cite{nist_data_base}. 
For these molecules, we investigated the relationship between the number of orbitals and the quantum resources 
by considering several Gaussian-type orbital basis sets for the smaller molecules (STO-6G, 6-31G*, cc-pVDZ, and cc-pVTZ).
All orbitals were included in the active space.
These small molecule benchmarks represent some of the most commonly found molecules, while stressing different types of molecular bonding. The choice of basis sets follows a progression of complexity, starting from where Gaussian orbitals are fit to a single Slater orbital (STO-6G)~\cite{stewart1970}, to more complete basis functions (6-31G*)~\cite{ditchfield1971} to more consistent basis sets designed for converging post-Hartree-Fock calculations (cc-pVDZ, cc-pVTZ)~\cite{dunning1989} to the complete basis set limit.

We also consider the much larger 
FeMoco molecule, also known as the iron-molybdenum cofactor, which is crucial for biological nitrogen fixation; however, its fixation mechanism is not 
fully understood \cite{reiher2017elucidating}. FeMoco is a well-known benchmark molecule that has been used in previous resource estimations. 
To compare our results with those of other studies, we used the same Hamiltonians (i.e., active spaces) 
as used in \cite{reiher2017elucidating} and \cite{li2019electronic}. 

\subsection{Resource Estimates}
\begin{figure}[!ht]
\begin{center}
\includegraphics[width=0.5\textwidth]{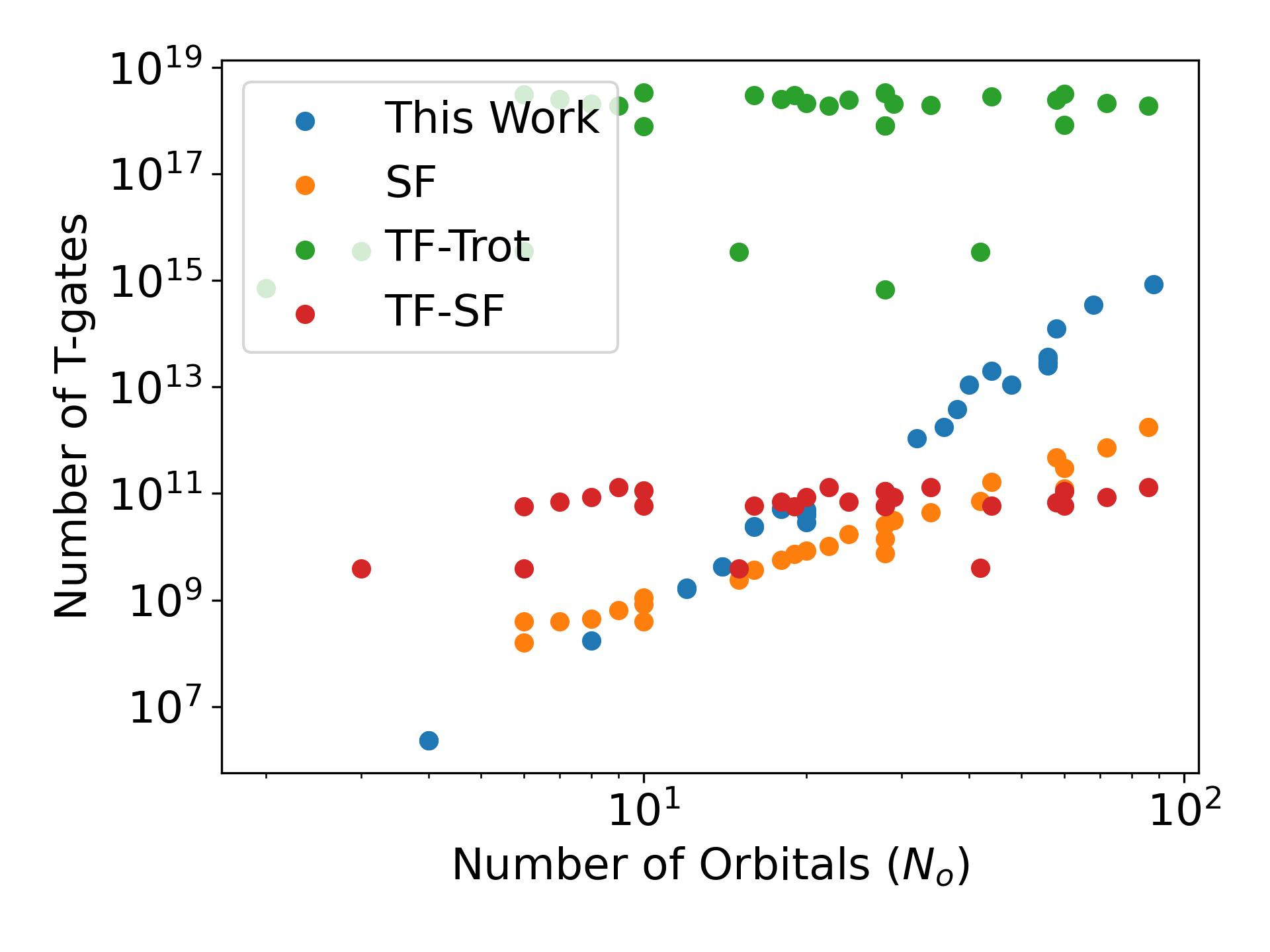}
\end{center}
\caption{Estimated total numbers of logical} T gates for various algorithms over many molecules at many basis set levels. See text for the definitions of the algorithms.\label{fig:resource}
\end{figure}

Using QREChem, we obtained the total number of logical T gates necessary for various algorithms for each molecule at different basis set levels. The data for all molecules is shown in \autoref{fig:resource}. 
``This Work" represents the QREChem estimate of number of Trotter gates, as described above. ``SF" uses OpenFermion's resource estimation tools for the single factorization algorithm~\cite{berry2019qubitization}
to estimate the number of Toffoli gates, which is then scaled by 4, as that is the number of T gates necessary for one Toffoli gate~\cite{berry2019qubitization}. ``TF-Trot" uses TFermion~\cite{casares2022tfermion} 
to estimate the number of T gates in the qDRIFT algorithm~\cite{campbell2019random} and ``TF-SF" uses TFermion to estimate the single factorization algorithm. Most striking is the comparison between
TFermion's Trotter algorithm (``TF-Trot") and the one estimated in QREChem (``This Work"). This is likely due to the fact the qDRIFT estimates use worst-case error bounds to calculate the total number
of operations. In QREChem's Trotter estimates, we instead use heuristic estimates, which result in orders of magnitude lower number of necessary Trotter steps
and, thus, orders of magnitude lower resource estimates. Furthermore, QREChem uses explicit gate counting
provided by the Microsoft QDK, rather than the more pessimistic estimates used in the formula of TFermion. These more realistic estimates put Trotter close to the estimates
of the more advanced single factorization algorithm
(which requires additional ancilla qubits).

We further compare the resource estimates of the number of logical T gates for various algorithms using FeMoco, which is the standard benchmark molecule for evaluation of quantum algorithms. We used the same Hamiltonian as in previous studies~\cite{reiher2017elucidating} and ~\cite{li2019electronic} to obtain a precise comparison with other works. \autoref{t7:FeMoco} shows the comparison of the number of T gates between QREChem's Trotter estimation
(``This Work") and others from \cite{casares2022tfermion,reiher2017elucidating,berry2019qubitization}. Ref.~\cite{reiher2017elucidating} provided the first theoretical estimates for FeMoco, which were approximately $10^{15}$ T gates. Again, we see that QREChem's Trotter estimation lines up with the estimation from Ref.~\cite{reiher2017elucidating}, which is reasonable, given that both works used similar heuristic estimates. The single factorization algorithm~\cite{berry2019qubitization} performs the best out of those studied. 

\begin{figure}[!ht]
\begin{center}
\includegraphics[width=0.4\textwidth]{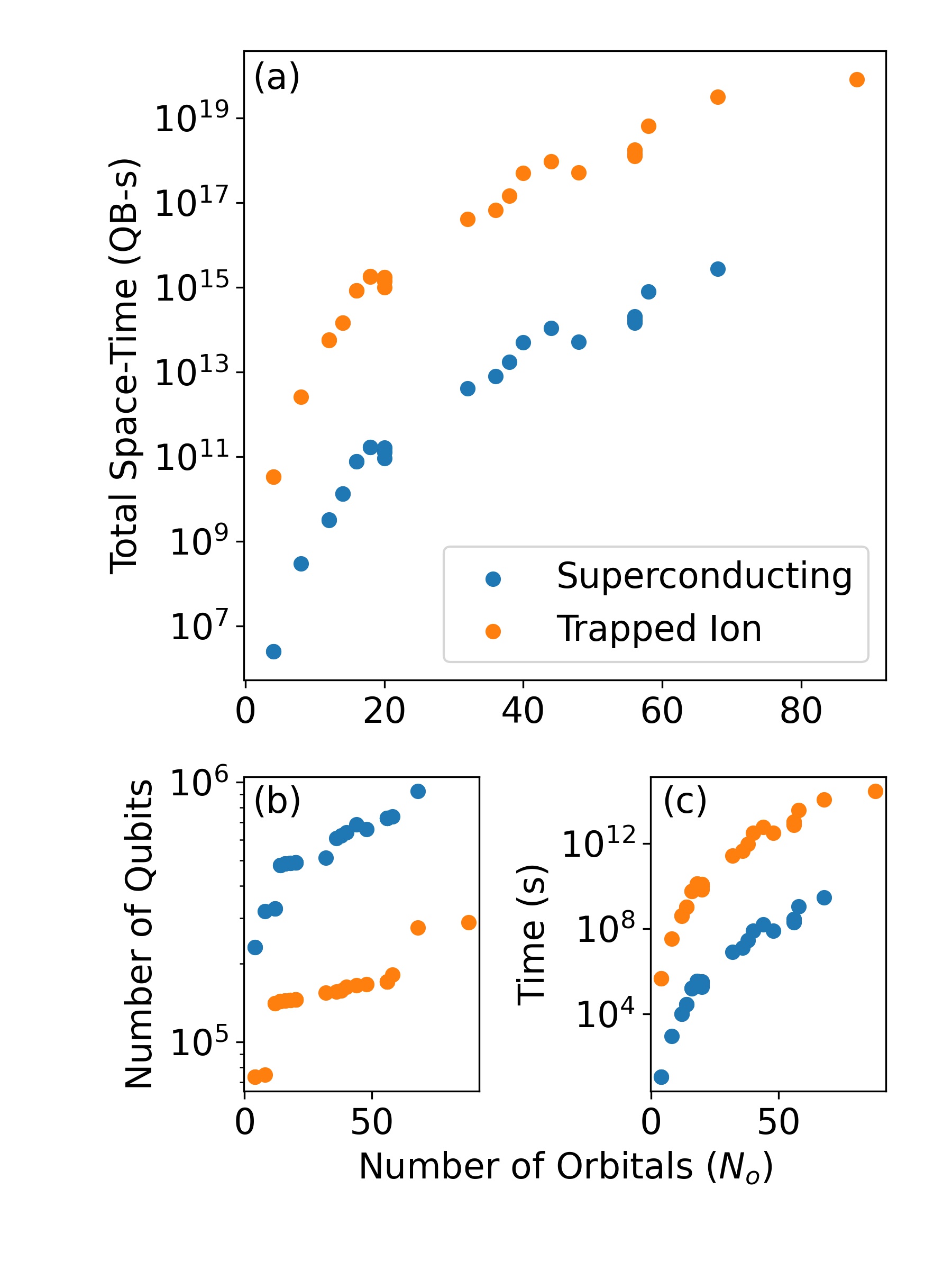}
\end{center}
\caption{Estimated total resources with hardware and surface code error correction overheads included for QREChem's Trotter algorithm. The total space-time volume (shown in (a)), in qubit-seconds, is larger for a trapped ion system compared with a superconducting qubit system. While the number of physical qubits is smaller for a trapped ion system due to the lower error rates (see (b)), the total time (see (c)) is much higher due to the slower error correction cycle time.}\label{fig:ec_overhead}
\end{figure}

All estimates presented in \autoref{fig:resource} and \autoref{t7:FeMoco}are only for the number of logical T gates required, not including additional overheads from hardware or error correction. 
To provide more realistic estimates, we estimated the surface code overhead for two benchmark systems: a 
superconducting qubit system and a trapped ion system. To parameterize the hardware, we use slight variations of 
parameters from recent demonstrations of error correction in each system~\cite{chen2021exponential,ryan2021realization}, 
which are summarized in \autoref{t7:hw_params}. The results are plotted in \autoref{fig:ec_overhead} for our benchmark small molecules. The much lower error 
rates of the trapped ion system used (3e-5 vs 5e-4 for the superconducting qubit system) allow for smaller surface code distances, $d$, and, hence, smaller numbers of physical qubits. The total runtime is several orders of magnitude higher, due to the increased cycle time (70ms vs 1$\mu$s for the superconducting qubit system). This leads to several orders of magnitude increase in the total space-time (measured in qubit-seconds) of the algorithm. These results point to the need for fast cycle times to achieve reasonable runtimes for quantum chemistry algorithms on quantum computers. Tabulated data for both physical qubit count and total runtime can be found in the Supplementary Materials for both architectures.

The data used in \autoref{fig:resource} in collated in tabular form in the Supplementary Materials. Further tables, including our parameter settings, the number of rotation and CNOT
gates and the number of physical qubits and runtime assuming a surface code error correction scheme on both superconducting qubit and trapped ion hardware can also be found in the Supplementary Materials as can details about the code and how to reproduce the results. The code is available from https://github.com/Argonne-QIS/QREChem/.




\section{Discussion}
We utilized the QREChem to provide accurate logical and physical resource estimations of a simple Trotter based algorithm over a broad range of molecules.
Our heuristic-based Trotter estimates offers a more realistic estimate of the true cost of using Trotter, compared
with more pessimistic estimates based on worst case bounds. On large systems such as the FeMoco molecule, the resource estimates are still large and do not include the 
necessary overheads of initial state preparation. The initial estimates of the overheads of quantum error correction and hardware limitations are significant and point to the need to have fast operations.

Accurate logical resource estimates, as currently implemented in QREChem, is a necessary first step for the larger goal of co-desiging future fault-tolerant quantum quantum computers 
capable of executing high-depth quantum chemistry circuits. 
Co-design allows for the optimization of both the hardware and algorithmic aspects of future quantum computers to facilitate the optimization of performance, scalability, accuracy, 
and energy efficiency. Moreover, co-design using a future version of QREChem will ensure that the simulation algorithms are tailored to the specific characteristics of the target quantum computers, ultimately 
enabling more effective and realistic quantum simulations.  Future developments will include a detailed examination and implementation of advanced features 
in the Hardware and Error Correction modules, as well as inclusion of additional quantum algorithms. These enhancements will allow QREChem to continue to provide accurate, comprehensive,
and actionable resource estimates for quantum chemistry.

\section*{Conflict of Interest Statement}
The authors declare that the research was conducted in the absence of any commercial or financial relationships that could be construed as a potential conflict of interest.

\section*{Author Contributions}

M.O., B.K., and D.F. performed the simulations. M.O., B.K., D.F, and Y.A. contributed to the code. All authors contributed important ideas during initial discussions and contributed to writing the manuscript.

\section*{Acknowledgments}
 This research used resources of the Argonne Leadership Computing Facility, which is a U.S. Department of Energy Office of Science User Facility operated under contract DE-AC02-06CH11357. This material is based upon work supported by the Defense Advanced Research Projects Agency under Contract No. HR001122C0074. Any opinions, findings and conclusions or recommendations expressed in this material are those of the author(s) and do not necessarily reflect the views of the Defense Advanced Research Projects Agency. This research was supported by the quantum computing technology development program of the National Research Foundation of Korea (NRF) funded by the Korean government (Ministry of Science and ICT(MSIT)) (No.2021M3H3A103657313) and by the National Research Foundation of Korea (NRF) grant funded by the Korea government(MOE) (No. 2020R1I1A2074957). This research is based on work supported by Laboratory Directed Research and Development (LDRD) funding from Argonne National Laboratory, provided by the Director, Office of Science, of the U.S. Department of Energy under Contract no. DE-AC0206CH11357. Work performed at the Center for Nanoscale Materials, a U.S. Department of Energy Office of Science User Facility, was supported by the U.S. DOE, Office of Basic Energy Sciences, under Contract No. DE-AC02-06CH11357.




\bibliographystyle{apsrev4-1}

\bibliography{main}

\begin{thebibliography}{55}%
\makeatletter
\providecommand \@ifxundefined [1]{%
 \@ifx{#1\undefined}
}%
\providecommand \@ifnum [1]{%
 \ifnum #1\expandafter \@firstoftwo
 \else \expandafter \@secondoftwo
 \fi
}%
\providecommand \@ifx [1]{%
 \ifx #1\expandafter \@firstoftwo
 \else \expandafter \@secondoftwo
 \fi
}%
\providecommand \natexlab [1]{#1}%
\providecommand \enquote  [1]{``#1''}%
\providecommand \bibnamefont  [1]{#1}%
\providecommand \bibfnamefont [1]{#1}%
\providecommand \citenamefont [1]{#1}%
\providecommand \href@noop [0]{\@secondoftwo}%
\providecommand \href [0]{\begingroup \@sanitize@url \@href}%
\providecommand \@href[1]{\@@startlink{#1}\@@href}%
\providecommand \@@href[1]{\endgroup#1\@@endlink}%
\providecommand \@sanitize@url [0]{\catcode `\\12\catcode `\$12\catcode `\&12\catcode `\#12\catcode `\^12\catcode `\_12\catcode `\%12\relax}%
\providecommand \@@startlink[1]{}%
\providecommand \@@endlink[0]{}%
\providecommand \url  [0]{\begingroup\@sanitize@url \@url }%
\providecommand \@url [1]{\endgroup\@href {#1}{\urlprefix }}%
\providecommand \urlprefix  [0]{URL }%
\providecommand \Eprint [0]{\href }%
\providecommand \doibase [0]{http://dx.doi.org/}%
\providecommand \selectlanguage [0]{\@gobble}%
\providecommand \bibinfo  [0]{\@secondoftwo}%
\providecommand \bibfield  [0]{\@secondoftwo}%
\providecommand \translation [1]{[#1]}%
\providecommand \BibitemOpen [0]{}%
\providecommand \bibitemStop [0]{}%
\providecommand \bibitemNoStop [0]{.\EOS\space}%
\providecommand \EOS [0]{\spacefactor3000\relax}%
\providecommand \BibitemShut  [1]{\csname bibitem#1\endcsname}%
\let\auto@bib@innerbib\@empty
\bibitem [{\citenamefont {Reiher}\ \emph {et~al.}(2017)\citenamefont {Reiher}, \citenamefont {Wiebe}, \citenamefont {Svore}, \citenamefont {Wecker},\ and\ \citenamefont {Troyer}}]{reiher2017elucidating}%
  \BibitemOpen
  \bibfield  {author} {\bibinfo {author} {\bibfnamefont {M.}~\bibnamefont {Reiher}}, \bibinfo {author} {\bibfnamefont {N.}~\bibnamefont {Wiebe}}, \bibinfo {author} {\bibfnamefont {K.~M.}\ \bibnamefont {Svore}}, \bibinfo {author} {\bibfnamefont {D.}~\bibnamefont {Wecker}}, \ and\ \bibinfo {author} {\bibfnamefont {M.}~\bibnamefont {Troyer}},\ }\href@noop {} {\bibfield  {journal} {\bibinfo  {journal} {Proceedings of the {N}ational {A}cademy of {S}ciences}\ }\textbf {\bibinfo {volume} {114}},\ \bibinfo {pages} {7555} (\bibinfo {year} {2017})}\BibitemShut {NoStop}%
\bibitem [{\citenamefont {Abrams}\ and\ \citenamefont {Lloyd}(1999)}]{abrams_quantum_1999}%
  \BibitemOpen
  \bibfield  {author} {\bibinfo {author} {\bibfnamefont {D.~S.}\ \bibnamefont {Abrams}}\ and\ \bibinfo {author} {\bibfnamefont {S.}~\bibnamefont {Lloyd}},\ }\href {\doibase 10.1103/PhysRevLett.83.5162} {\bibfield  {journal} {\bibinfo  {journal} {Physical Review Letters}\ }\textbf {\bibinfo {volume} {83}},\ \bibinfo {pages} {5162} (\bibinfo {year} {1999})}\BibitemShut {NoStop}%
\bibitem [{\citenamefont {Kitaev}(1997)}]{kitaev_quantum_1997}%
  \BibitemOpen
  \bibfield  {author} {\bibinfo {author} {\bibfnamefont {A.~Y.}\ \bibnamefont {Kitaev}},\ }\href {\doibase 10.1070/RM1997v052n06ABEH002155} {\bibfield  {journal} {\bibinfo  {journal} {Russian Mathematical Surveys}\ }\textbf {\bibinfo {volume} {52}},\ \bibinfo {pages} {1191} (\bibinfo {year} {1997})}\BibitemShut {NoStop}%
\bibitem [{\citenamefont {Kim}\ \emph {et~al.}(2022)\citenamefont {Kim}, \citenamefont {Liu}, \citenamefont {Pallister}, \citenamefont {Pol}, \citenamefont {Roberts},\ and\ \citenamefont {Lee}}]{kim2022fault}%
  \BibitemOpen
  \bibfield  {author} {\bibinfo {author} {\bibfnamefont {I.~H.}\ \bibnamefont {Kim}}, \bibinfo {author} {\bibfnamefont {Y.-H.}\ \bibnamefont {Liu}}, \bibinfo {author} {\bibfnamefont {S.}~\bibnamefont {Pallister}}, \bibinfo {author} {\bibfnamefont {W.}~\bibnamefont {Pol}}, \bibinfo {author} {\bibfnamefont {S.}~\bibnamefont {Roberts}}, \ and\ \bibinfo {author} {\bibfnamefont {E.}~\bibnamefont {Lee}},\ }\href@noop {} {\bibfield  {journal} {\bibinfo  {journal} {Physical Review Research}\ }\textbf {\bibinfo {volume} {4}},\ \bibinfo {pages} {023019} (\bibinfo {year} {2022})}\BibitemShut {NoStop}%
\bibitem [{\citenamefont {Preskill}(2018)}]{preskill2018quantum}%
  \BibitemOpen
  \bibfield  {author} {\bibinfo {author} {\bibfnamefont {J.}~\bibnamefont {Preskill}},\ }\href@noop {} {\bibfield  {journal} {\bibinfo  {journal} {Quantum}\ }\textbf {\bibinfo {volume} {2}},\ \bibinfo {pages} {79} (\bibinfo {year} {2018})}\BibitemShut {NoStop}%
\bibitem [{\citenamefont {Peruzzo}\ \emph {et~al.}(2014)\citenamefont {Peruzzo}, \citenamefont {McClean}, \citenamefont {Shadbolt}, \citenamefont {Yung}, \citenamefont {Zhou}, \citenamefont {Love}, \citenamefont {Aspuru-Guzik},\ and\ \citenamefont {O'Brien}}]{Peruzzo2014}%
  \BibitemOpen
  \bibfield  {author} {\bibinfo {author} {\bibfnamefont {A.}~\bibnamefont {Peruzzo}}, \bibinfo {author} {\bibfnamefont {J.}~\bibnamefont {McClean}}, \bibinfo {author} {\bibfnamefont {P.}~\bibnamefont {Shadbolt}}, \bibinfo {author} {\bibfnamefont {M.~H.}\ \bibnamefont {Yung}}, \bibinfo {author} {\bibfnamefont {X.~Q.}\ \bibnamefont {Zhou}}, \bibinfo {author} {\bibfnamefont {P.~J.}\ \bibnamefont {Love}}, \bibinfo {author} {\bibfnamefont {A.}~\bibnamefont {Aspuru-Guzik}}, \ and\ \bibinfo {author} {\bibfnamefont {J.~L.}\ \bibnamefont {O'Brien}},\ }\href {\doibase 10.1038/ncomms5213} {\bibfield  {journal} {\bibinfo  {journal} {Nature Communications}\ }\textbf {\bibinfo {volume} {5}},\ \bibinfo {pages} {4213} (\bibinfo {year} {2014})}\BibitemShut {NoStop}%
\bibitem [{\citenamefont {Menickelly}\ \emph {et~al.}(2023)\citenamefont {Menickelly}, \citenamefont {Ha},\ and\ \citenamefont {Otten}}]{Menickelly2023latency}%
  \BibitemOpen
  \bibfield  {author} {\bibinfo {author} {\bibfnamefont {M.}~\bibnamefont {Menickelly}}, \bibinfo {author} {\bibfnamefont {Y.}~\bibnamefont {Ha}}, \ and\ \bibinfo {author} {\bibfnamefont {M.}~\bibnamefont {Otten}},\ }\href {\doibase 10.22331/q-2023-03-16-949} {\bibfield  {journal} {\bibinfo  {journal} {{Quantum}}\ }\textbf {\bibinfo {volume} {7}},\ \bibinfo {pages} {949} (\bibinfo {year} {2023})}\BibitemShut {NoStop}%
\bibitem [{\citenamefont {Casares}\ \emph {et~al.}(2022)\citenamefont {Casares}, \citenamefont {Campos},\ and\ \citenamefont {Martin-Delgado}}]{casares2022tfermion}%
  \BibitemOpen
  \bibfield  {author} {\bibinfo {author} {\bibfnamefont {P.~A.}\ \bibnamefont {Casares}}, \bibinfo {author} {\bibfnamefont {R.}~\bibnamefont {Campos}}, \ and\ \bibinfo {author} {\bibfnamefont {M.~A.}\ \bibnamefont {Martin-Delgado}},\ }\href@noop {} {\bibfield  {journal} {\bibinfo  {journal} {Quantum}\ }\textbf {\bibinfo {volume} {6}},\ \bibinfo {pages} {768} (\bibinfo {year} {2022})}\BibitemShut {NoStop}%
\bibitem [{\citenamefont {McClean}\ \emph {et~al.}(2020)\citenamefont {McClean}, \citenamefont {Rubin}, \citenamefont {Sung}, \citenamefont {Kivlichan}, \citenamefont {Bonet-Monroig}, \citenamefont {Cao}, \citenamefont {Dai}, \citenamefont {Fried}, \citenamefont {Gidney}, \citenamefont {Gimby} \emph {et~al.}}]{mcclean2020openfermion}%
  \BibitemOpen
  \bibfield  {author} {\bibinfo {author} {\bibfnamefont {J.~R.}\ \bibnamefont {McClean}}, \bibinfo {author} {\bibfnamefont {N.~C.}\ \bibnamefont {Rubin}}, \bibinfo {author} {\bibfnamefont {K.~J.}\ \bibnamefont {Sung}}, \bibinfo {author} {\bibfnamefont {I.~D.}\ \bibnamefont {Kivlichan}}, \bibinfo {author} {\bibfnamefont {X.}~\bibnamefont {Bonet-Monroig}}, \bibinfo {author} {\bibfnamefont {Y.}~\bibnamefont {Cao}}, \bibinfo {author} {\bibfnamefont {C.}~\bibnamefont {Dai}}, \bibinfo {author} {\bibfnamefont {E.~S.}\ \bibnamefont {Fried}}, \bibinfo {author} {\bibfnamefont {C.}~\bibnamefont {Gidney}}, \bibinfo {author} {\bibfnamefont {B.}~\bibnamefont {Gimby}},  \emph {et~al.},\ }\href@noop {} {\bibfield  {journal} {\bibinfo  {journal} {Quantum Science and Technology}\ }\textbf {\bibinfo {volume} {5}},\ \bibinfo {pages} {034014} (\bibinfo {year} {2020})}\BibitemShut {NoStop}%
\bibitem [{\citenamefont {Beverland}\ \emph {et~al.}(2022)\citenamefont {Beverland}, \citenamefont {Murali}, \citenamefont {Troyer}, \citenamefont {Svore}, \citenamefont {Hoeffler}, \citenamefont {Kliuchnikov}, \citenamefont {Low}, \citenamefont {Soeken}, \citenamefont {Sundaram},\ and\ \citenamefont {Vaschillo}}]{beverland2022assessing}%
  \BibitemOpen
  \bibfield  {author} {\bibinfo {author} {\bibfnamefont {M.~E.}\ \bibnamefont {Beverland}}, \bibinfo {author} {\bibfnamefont {P.}~\bibnamefont {Murali}}, \bibinfo {author} {\bibfnamefont {M.}~\bibnamefont {Troyer}}, \bibinfo {author} {\bibfnamefont {K.~M.}\ \bibnamefont {Svore}}, \bibinfo {author} {\bibfnamefont {T.}~\bibnamefont {Hoeffler}}, \bibinfo {author} {\bibfnamefont {V.}~\bibnamefont {Kliuchnikov}}, \bibinfo {author} {\bibfnamefont {G.~H.}\ \bibnamefont {Low}}, \bibinfo {author} {\bibfnamefont {M.}~\bibnamefont {Soeken}}, \bibinfo {author} {\bibfnamefont {A.}~\bibnamefont {Sundaram}}, \ and\ \bibinfo {author} {\bibfnamefont {A.}~\bibnamefont {Vaschillo}},\ }\href@noop {} {\bibfield  {journal} {\bibinfo  {journal} {arXiv preprint arXiv:2211.07629}\ } (\bibinfo {year} {2022})}\BibitemShut {NoStop}%
\bibitem [{\citenamefont {Ortiz}\ \emph {et~al.}(2001)\citenamefont {Ortiz}, \citenamefont {Gubernatis}, \citenamefont {Knill},\ and\ \citenamefont {Laflamme}}]{PhysRevA.64.022319}%
  \BibitemOpen
  \bibfield  {author} {\bibinfo {author} {\bibfnamefont {G.}~\bibnamefont {Ortiz}}, \bibinfo {author} {\bibfnamefont {J.~E.}\ \bibnamefont {Gubernatis}}, \bibinfo {author} {\bibfnamefont {E.}~\bibnamefont {Knill}}, \ and\ \bibinfo {author} {\bibfnamefont {R.}~\bibnamefont {Laflamme}},\ }\href {\doibase 10.1103/PhysRevA.64.022319} {\bibfield  {journal} {\bibinfo  {journal} {Phys. Rev. A}\ }\textbf {\bibinfo {volume} {64}},\ \bibinfo {pages} {022319} (\bibinfo {year} {2001})}\BibitemShut {NoStop}%
\bibitem [{\citenamefont {Babbush}\ \emph {et~al.}(2015)\citenamefont {Babbush}, \citenamefont {McClean}, \citenamefont {Wecker}, \citenamefont {Aspuru-Guzik},\ and\ \citenamefont {Wiebe}}]{PhysRevA.91.022311}%
  \BibitemOpen
  \bibfield  {author} {\bibinfo {author} {\bibfnamefont {R.}~\bibnamefont {Babbush}}, \bibinfo {author} {\bibfnamefont {J.}~\bibnamefont {McClean}}, \bibinfo {author} {\bibfnamefont {D.}~\bibnamefont {Wecker}}, \bibinfo {author} {\bibfnamefont {A.}~\bibnamefont {Aspuru-Guzik}}, \ and\ \bibinfo {author} {\bibfnamefont {N.}~\bibnamefont {Wiebe}},\ }\href {\doibase 10.1103/PhysRevA.91.022311} {\bibfield  {journal} {\bibinfo  {journal} {Phys. Rev. A}\ }\textbf {\bibinfo {volume} {91}},\ \bibinfo {pages} {022311} (\bibinfo {year} {2015})}\BibitemShut {NoStop}%
\bibitem [{\citenamefont {Sun}\ \emph {et~al.}(2018)\citenamefont {Sun}, \citenamefont {Berkelbach}, \citenamefont {Blunt}, \citenamefont {Booth}, \citenamefont {Guo}, \citenamefont {Li}, \citenamefont {Liu}, \citenamefont {McClain}, \citenamefont {Sayfutyarova}, \citenamefont {Sharma} \emph {et~al.}}]{sun2018pyscf}%
  \BibitemOpen
  \bibfield  {author} {\bibinfo {author} {\bibfnamefont {Q.}~\bibnamefont {Sun}}, \bibinfo {author} {\bibfnamefont {T.~C.}\ \bibnamefont {Berkelbach}}, \bibinfo {author} {\bibfnamefont {N.~S.}\ \bibnamefont {Blunt}}, \bibinfo {author} {\bibfnamefont {G.~H.}\ \bibnamefont {Booth}}, \bibinfo {author} {\bibfnamefont {S.}~\bibnamefont {Guo}}, \bibinfo {author} {\bibfnamefont {Z.}~\bibnamefont {Li}}, \bibinfo {author} {\bibfnamefont {J.}~\bibnamefont {Liu}}, \bibinfo {author} {\bibfnamefont {J.~D.}\ \bibnamefont {McClain}}, \bibinfo {author} {\bibfnamefont {E.~R.}\ \bibnamefont {Sayfutyarova}}, \bibinfo {author} {\bibfnamefont {S.}~\bibnamefont {Sharma}},  \emph {et~al.},\ }\href@noop {} {\bibfield  {journal} {\bibinfo  {journal} {Wiley Interdisciplinary Reviews: Computational Molecular Science}\ }\textbf {\bibinfo {volume} {8}},\ \bibinfo {pages} {e1340} (\bibinfo {year} {2018})}\BibitemShut {NoStop}%
\bibitem [{\citenamefont {Sun}\ \emph {et~al.}(2020)\citenamefont {Sun}, \citenamefont {Zhang}, \citenamefont {Banerjee}, \citenamefont {Bao}, \citenamefont {Barbry}, \citenamefont {Blunt}, \citenamefont {Bogdanov}, \citenamefont {Booth}, \citenamefont {Chen}, \citenamefont {Cui} \emph {et~al.}}]{sun2020recent}%
  \BibitemOpen
  \bibfield  {author} {\bibinfo {author} {\bibfnamefont {Q.}~\bibnamefont {Sun}}, \bibinfo {author} {\bibfnamefont {X.}~\bibnamefont {Zhang}}, \bibinfo {author} {\bibfnamefont {S.}~\bibnamefont {Banerjee}}, \bibinfo {author} {\bibfnamefont {P.}~\bibnamefont {Bao}}, \bibinfo {author} {\bibfnamefont {M.}~\bibnamefont {Barbry}}, \bibinfo {author} {\bibfnamefont {N.~S.}\ \bibnamefont {Blunt}}, \bibinfo {author} {\bibfnamefont {N.~A.}\ \bibnamefont {Bogdanov}}, \bibinfo {author} {\bibfnamefont {G.~H.}\ \bibnamefont {Booth}}, \bibinfo {author} {\bibfnamefont {J.}~\bibnamefont {Chen}}, \bibinfo {author} {\bibfnamefont {Z.-H.}\ \bibnamefont {Cui}},  \emph {et~al.},\ }\href@noop {} {\bibfield  {journal} {\bibinfo  {journal} {The Journal of Chemical Physics}\ }\textbf {\bibinfo {volume} {153}},\ \bibinfo {pages} {024109} (\bibinfo {year} {2020})}\BibitemShut {NoStop}%
\bibitem [{\citenamefont {Low}\ \emph {et~al.}(2019)\citenamefont {Low}, \citenamefont {Bauman}, \citenamefont {Granade}, \citenamefont {Peng}, \citenamefont {Wiebe}, \citenamefont {Bylaska}, \citenamefont {Wecker}, \citenamefont {Krishnamoorthy}, \citenamefont {Roetteler}, \citenamefont {Kowalski} \emph {et~al.}}]{low2019q}%
  \BibitemOpen
  \bibfield  {author} {\bibinfo {author} {\bibfnamefont {G.~H.}\ \bibnamefont {Low}}, \bibinfo {author} {\bibfnamefont {N.~P.}\ \bibnamefont {Bauman}}, \bibinfo {author} {\bibfnamefont {C.~E.}\ \bibnamefont {Granade}}, \bibinfo {author} {\bibfnamefont {B.}~\bibnamefont {Peng}}, \bibinfo {author} {\bibfnamefont {N.}~\bibnamefont {Wiebe}}, \bibinfo {author} {\bibfnamefont {E.~J.}\ \bibnamefont {Bylaska}}, \bibinfo {author} {\bibfnamefont {D.}~\bibnamefont {Wecker}}, \bibinfo {author} {\bibfnamefont {S.}~\bibnamefont {Krishnamoorthy}}, \bibinfo {author} {\bibfnamefont {M.}~\bibnamefont {Roetteler}}, \bibinfo {author} {\bibfnamefont {K.}~\bibnamefont {Kowalski}},  \emph {et~al.},\ }\href@noop {} {\bibfield  {journal} {\bibinfo  {journal} {arXiv preprint arXiv:1904.01131}\ } (\bibinfo {year} {2019})}\BibitemShut {NoStop}%
\bibitem [{\citenamefont {Knowles}\ and\ \citenamefont {Handy}(1989)}]{knowles1989determinant}%
  \BibitemOpen
  \bibfield  {author} {\bibinfo {author} {\bibfnamefont {P.~J.}\ \bibnamefont {Knowles}}\ and\ \bibinfo {author} {\bibfnamefont {N.~C.}\ \bibnamefont {Handy}},\ }\href@noop {} {\bibfield  {journal} {\bibinfo  {journal} {Computer {P}hysics {C}ommunications}\ }\textbf {\bibinfo {volume} {54}},\ \bibinfo {pages} {75} (\bibinfo {year} {1989})}\BibitemShut {NoStop}%
\bibitem [{\citenamefont {Foresman}\ and\ \citenamefont {Frish}(1996)}]{foresman1996exploring}%
  \BibitemOpen
  \bibfield  {author} {\bibinfo {author} {\bibfnamefont {J.}~\bibnamefont {Foresman}}\ and\ \bibinfo {author} {\bibfnamefont {E.}~\bibnamefont {Frish}},\ }\href@noop {} {\bibfield  {journal} {\bibinfo  {journal} {Gaussian Inc., Pittsburg, USA}\ }\textbf {\bibinfo {volume} {21}} (\bibinfo {year} {1996})}\BibitemShut {NoStop}%
\bibitem [{\citenamefont {Werner}\ \emph {et~al.}(2012)\citenamefont {Werner}, \citenamefont {Knowles}, \citenamefont {Knizia}, \citenamefont {Manby},\ and\ \citenamefont {Sch{\"u}tz}}]{werner2012molpro}%
  \BibitemOpen
  \bibfield  {author} {\bibinfo {author} {\bibfnamefont {H.-J.}\ \bibnamefont {Werner}}, \bibinfo {author} {\bibfnamefont {P.~J.}\ \bibnamefont {Knowles}}, \bibinfo {author} {\bibfnamefont {G.}~\bibnamefont {Knizia}}, \bibinfo {author} {\bibfnamefont {F.~R.}\ \bibnamefont {Manby}}, \ and\ \bibinfo {author} {\bibfnamefont {M.}~\bibnamefont {Sch{\"u}tz}},\ }\href@noop {} {\bibfield  {journal} {\bibinfo  {journal} {Wiley Interdisciplinary Reviews: Computational Molecular Science}\ }\textbf {\bibinfo {volume} {2}},\ \bibinfo {pages} {242} (\bibinfo {year} {2012})}\BibitemShut {NoStop}%
\bibitem [{\citenamefont {Smith}\ \emph {et~al.}(2020)\citenamefont {Smith}, \citenamefont {Burns}, \citenamefont {Simmonett}, \citenamefont {Parrish}, \citenamefont {Schieber}, \citenamefont {Galvelis}, \citenamefont {Kraus}, \citenamefont {Kruse}, \citenamefont {Di~Remigio}, \citenamefont {Alenaizan} \emph {et~al.}}]{smith2020psi4}%
  \BibitemOpen
  \bibfield  {author} {\bibinfo {author} {\bibfnamefont {D.~G.}\ \bibnamefont {Smith}}, \bibinfo {author} {\bibfnamefont {L.~A.}\ \bibnamefont {Burns}}, \bibinfo {author} {\bibfnamefont {A.~C.}\ \bibnamefont {Simmonett}}, \bibinfo {author} {\bibfnamefont {R.~M.}\ \bibnamefont {Parrish}}, \bibinfo {author} {\bibfnamefont {M.~C.}\ \bibnamefont {Schieber}}, \bibinfo {author} {\bibfnamefont {R.}~\bibnamefont {Galvelis}}, \bibinfo {author} {\bibfnamefont {P.}~\bibnamefont {Kraus}}, \bibinfo {author} {\bibfnamefont {H.}~\bibnamefont {Kruse}}, \bibinfo {author} {\bibfnamefont {R.}~\bibnamefont {Di~Remigio}}, \bibinfo {author} {\bibfnamefont {A.}~\bibnamefont {Alenaizan}},  \emph {et~al.},\ }\href@noop {} {\bibfield  {journal} {\bibinfo  {journal} {The Journal of chemical physics}\ }\textbf {\bibinfo {volume} {152}} (\bibinfo {year} {2020})}\BibitemShut {NoStop}%
\bibitem [{\citenamefont {Otten}\ \emph {et~al.}(2022)\citenamefont {Otten}, \citenamefont {Hermes}, \citenamefont {Pandharkar}, \citenamefont {Alexeev}, \citenamefont {Gray},\ and\ \citenamefont {Gagliardi}}]{otten2022localized}%
  \BibitemOpen
  \bibfield  {author} {\bibinfo {author} {\bibfnamefont {M.}~\bibnamefont {Otten}}, \bibinfo {author} {\bibfnamefont {M.~R.}\ \bibnamefont {Hermes}}, \bibinfo {author} {\bibfnamefont {R.}~\bibnamefont {Pandharkar}}, \bibinfo {author} {\bibfnamefont {Y.}~\bibnamefont {Alexeev}}, \bibinfo {author} {\bibfnamefont {S.~K.}\ \bibnamefont {Gray}}, \ and\ \bibinfo {author} {\bibfnamefont {L.}~\bibnamefont {Gagliardi}},\ }\href@noop {} {\bibfield  {journal} {\bibinfo  {journal} {Journal of Chemical Theory and Computation}\ }\textbf {\bibinfo {volume} {18}},\ \bibinfo {pages} {7205} (\bibinfo {year} {2022})}\BibitemShut {NoStop}%
\bibitem [{\citenamefont {D'Cunha}\ \emph {et~al.}(2023)\citenamefont {D'Cunha}, \citenamefont {Otten}, \citenamefont {Hermes}, \citenamefont {Gagliardi},\ and\ \citenamefont {Gray}}]{dcunha2023state}%
  \BibitemOpen
  \bibfield  {author} {\bibinfo {author} {\bibfnamefont {R.}~\bibnamefont {D'Cunha}}, \bibinfo {author} {\bibfnamefont {M.}~\bibnamefont {Otten}}, \bibinfo {author} {\bibfnamefont {M.~R.}\ \bibnamefont {Hermes}}, \bibinfo {author} {\bibfnamefont {L.}~\bibnamefont {Gagliardi}}, \ and\ \bibinfo {author} {\bibfnamefont {S.~K.}\ \bibnamefont {Gray}},\ }\href@noop {} {\bibfield  {journal} {\bibinfo  {journal} {arXiv preprint arXiv:2305.18110}\ } (\bibinfo {year} {2023})}\BibitemShut {NoStop}%
\bibitem [{\citenamefont {Xia}\ and\ \citenamefont {Kais}(2018)}]{xia2018quantum}%
  \BibitemOpen
  \bibfield  {author} {\bibinfo {author} {\bibfnamefont {R.}~\bibnamefont {Xia}}\ and\ \bibinfo {author} {\bibfnamefont {S.}~\bibnamefont {Kais}},\ }\href@noop {} {\bibfield  {journal} {\bibinfo  {journal} {Nature communications}\ }\textbf {\bibinfo {volume} {9}},\ \bibinfo {pages} {4195} (\bibinfo {year} {2018})}\BibitemShut {NoStop}%
\bibitem [{\citenamefont {Shor}(1999)}]{shor1999polynomial}%
  \BibitemOpen
  \bibfield  {author} {\bibinfo {author} {\bibfnamefont {P.~W.}\ \bibnamefont {Shor}},\ }\href {\doibase 10.1137/S0036144598347011} {\bibfield  {journal} {\bibinfo  {journal} {SIAM Review}\ }\textbf {\bibinfo {volume} {41}},\ \bibinfo {pages} {303} (\bibinfo {year} {1999})}\BibitemShut {NoStop}%
\bibitem [{\citenamefont {Harrow}\ \emph {et~al.}(2009)\citenamefont {Harrow}, \citenamefont {Hassidim},\ and\ \citenamefont {Lloyd}}]{PhysRevLett.103.150502}%
  \BibitemOpen
  \bibfield  {author} {\bibinfo {author} {\bibfnamefont {A.~W.}\ \bibnamefont {Harrow}}, \bibinfo {author} {\bibfnamefont {A.}~\bibnamefont {Hassidim}}, \ and\ \bibinfo {author} {\bibfnamefont {S.}~\bibnamefont {Lloyd}},\ }\href {\doibase 10.1103/PhysRevLett.103.150502} {\bibfield  {journal} {\bibinfo  {journal} {Phys. Rev. Lett.}\ }\textbf {\bibinfo {volume} {103}},\ \bibinfo {pages} {150502} (\bibinfo {year} {2009})}\BibitemShut {NoStop}%
\bibitem [{\citenamefont {Berry}\ \emph {et~al.}(2015)\citenamefont {Berry}, \citenamefont {Childs}, \citenamefont {Cleve}, \citenamefont {Kothari},\ and\ \citenamefont {Somma}}]{PhysRevLett.114.090502}%
  \BibitemOpen
  \bibfield  {author} {\bibinfo {author} {\bibfnamefont {D.~W.}\ \bibnamefont {Berry}}, \bibinfo {author} {\bibfnamefont {A.~M.}\ \bibnamefont {Childs}}, \bibinfo {author} {\bibfnamefont {R.}~\bibnamefont {Cleve}}, \bibinfo {author} {\bibfnamefont {R.}~\bibnamefont {Kothari}}, \ and\ \bibinfo {author} {\bibfnamefont {R.~D.}\ \bibnamefont {Somma}},\ }\href {\doibase 10.1103/PhysRevLett.114.090502} {\bibfield  {journal} {\bibinfo  {journal} {Phys. Rev. Lett.}\ }\textbf {\bibinfo {volume} {114}},\ \bibinfo {pages} {090502} (\bibinfo {year} {2015})}\BibitemShut {NoStop}%
\bibitem [{\citenamefont {Low}\ and\ \citenamefont {Chuang}(2019)}]{Low2019hamiltonian}%
  \BibitemOpen
  \bibfield  {author} {\bibinfo {author} {\bibfnamefont {G.~H.}\ \bibnamefont {Low}}\ and\ \bibinfo {author} {\bibfnamefont {I.~L.}\ \bibnamefont {Chuang}},\ }\href {\doibase 10.22331/q-2019-07-12-163} {\bibfield  {journal} {\bibinfo  {journal} {{Quantum}}\ }\textbf {\bibinfo {volume} {3}},\ \bibinfo {pages} {163} (\bibinfo {year} {2019})}\BibitemShut {NoStop}%
\bibitem [{\citenamefont {Suzuki}(1993)}]{suzuki1993improved}%
  \BibitemOpen
  \bibfield  {author} {\bibinfo {author} {\bibfnamefont {M.}~\bibnamefont {Suzuki}},\ }\href@noop {} {\bibfield  {journal} {\bibinfo  {journal} {Physics Letters A}\ }\textbf {\bibinfo {volume} {180}},\ \bibinfo {pages} {232} (\bibinfo {year} {1993})}\BibitemShut {NoStop}%
\bibitem [{\citenamefont {Jordan}\ and\ \citenamefont {Wigner}(1993)}]{jordan1993paulische}%
  \BibitemOpen
  \bibfield  {author} {\bibinfo {author} {\bibfnamefont {P.}~\bibnamefont {Jordan}}\ and\ \bibinfo {author} {\bibfnamefont {E.~P.}\ \bibnamefont {Wigner}},\ }in\ \href@noop {} {\emph {\bibinfo {booktitle} {The Collected Works of Eugene Paul Wigner}}}\ (\bibinfo  {publisher} {Springer},\ \bibinfo {year} {1993})\ pp.\ \bibinfo {pages} {109--129}\BibitemShut {NoStop}%
\bibitem [{\citenamefont {Shor}(1994)}]{365700}%
  \BibitemOpen
  \bibfield  {author} {\bibinfo {author} {\bibfnamefont {P.}~\bibnamefont {Shor}},\ }in\ \href {\doibase 10.1109/SFCS.1994.365700} {\emph {\bibinfo {booktitle} {Proceedings 35th Annual Symposium on Foundations of Computer Science}}}\ (\bibinfo {year} {1994})\ pp.\ \bibinfo {pages} {124--134}\BibitemShut {NoStop}%
\bibitem [{\citenamefont {Svore}\ \emph {et~al.}(2018)\citenamefont {Svore}, \citenamefont {Geller}, \citenamefont {Troyer}, \citenamefont {Azariah}, \citenamefont {Granade}, \citenamefont {Heim}, \citenamefont {Kliuchnikov}, \citenamefont {Mykhailova}, \citenamefont {Paz},\ and\ \citenamefont {Roetteler}}]{svore2018qsharp}%
  \BibitemOpen
  \bibfield  {author} {\bibinfo {author} {\bibfnamefont {K.}~\bibnamefont {Svore}}, \bibinfo {author} {\bibfnamefont {A.}~\bibnamefont {Geller}}, \bibinfo {author} {\bibfnamefont {M.}~\bibnamefont {Troyer}}, \bibinfo {author} {\bibfnamefont {J.}~\bibnamefont {Azariah}}, \bibinfo {author} {\bibfnamefont {C.}~\bibnamefont {Granade}}, \bibinfo {author} {\bibfnamefont {B.}~\bibnamefont {Heim}}, \bibinfo {author} {\bibfnamefont {V.}~\bibnamefont {Kliuchnikov}}, \bibinfo {author} {\bibfnamefont {M.}~\bibnamefont {Mykhailova}}, \bibinfo {author} {\bibfnamefont {A.}~\bibnamefont {Paz}}, \ and\ \bibinfo {author} {\bibfnamefont {M.}~\bibnamefont {Roetteler}},\ }in\ \href {\doibase 10.1145/3183895.3183901} {\emph {\bibinfo {booktitle} {Proceedings of the Real World Domain Specific Languages Workshop 2018}}},\ \bibinfo {series and number} {RWDSL2018}\ (\bibinfo  {publisher} {Association for Computing Machinery},\ \bibinfo {address} {New York, NY, USA},\ \bibinfo {year} {2018})\BibitemShut {NoStop}%
\bibitem [{\citenamefont {Poulin}\ \emph {et~al.}(2015)\citenamefont {Poulin}, \citenamefont {Hastings}, \citenamefont {Wecker}, \citenamefont {Wiebe}, \citenamefont {Doberty},\ and\ \citenamefont {Troyer}}]{poulin2015thetrotter}%
  \BibitemOpen
  \bibfield  {author} {\bibinfo {author} {\bibfnamefont {D.}~\bibnamefont {Poulin}}, \bibinfo {author} {\bibfnamefont {M.~B.}\ \bibnamefont {Hastings}}, \bibinfo {author} {\bibfnamefont {D.}~\bibnamefont {Wecker}}, \bibinfo {author} {\bibfnamefont {N.}~\bibnamefont {Wiebe}}, \bibinfo {author} {\bibfnamefont {A.~C.}\ \bibnamefont {Doberty}}, \ and\ \bibinfo {author} {\bibfnamefont {M.}~\bibnamefont {Troyer}},\ }\href@noop {} {\bibfield  {journal} {\bibinfo  {journal} {Quantum Info. Comput.}\ }\textbf {\bibinfo {volume} {15}},\ \bibinfo {pages} {361–384} (\bibinfo {year} {2015})}\BibitemShut {NoStop}%
\bibitem [{\citenamefont {Nielsen}\ and\ \citenamefont {Chuang}(2000)}]{nielsen2000quantum}%
  \BibitemOpen
  \bibfield  {author} {\bibinfo {author} {\bibfnamefont {M.~A.}\ \bibnamefont {Nielsen}}\ and\ \bibinfo {author} {\bibfnamefont {I.~L.}\ \bibnamefont {Chuang}},\ }\href@noop {} {\emph {\bibinfo {title} {Quantum Computation and Quantum Information}}}\ (\bibinfo  {publisher} {Cambridge University Press},\ \bibinfo {year} {2000})\BibitemShut {NoStop}%
\bibitem [{\citenamefont {Li}(2022)}]{li2022some}%
  \BibitemOpen
  \bibfield  {author} {\bibinfo {author} {\bibfnamefont {X.}~\bibnamefont {Li}},\ }\href@noop {} {\bibfield  {journal} {\bibinfo  {journal} {Journal of Physics A: Mathematical and Theoretical}\ } (\bibinfo {year} {2022})}\BibitemShut {NoStop}%
\bibitem [{\citenamefont {Chien}\ \emph {et~al.}(2018)\citenamefont {Chien}, \citenamefont {Holmes}, \citenamefont {Otten}, \citenamefont {Umrigar}, \citenamefont {Sharma},\ and\ \citenamefont {Zimmerman}}]{chien2018excited}%
  \BibitemOpen
  \bibfield  {author} {\bibinfo {author} {\bibfnamefont {A.~D.}\ \bibnamefont {Chien}}, \bibinfo {author} {\bibfnamefont {A.~A.}\ \bibnamefont {Holmes}}, \bibinfo {author} {\bibfnamefont {M.}~\bibnamefont {Otten}}, \bibinfo {author} {\bibfnamefont {C.~J.}\ \bibnamefont {Umrigar}}, \bibinfo {author} {\bibfnamefont {S.}~\bibnamefont {Sharma}}, \ and\ \bibinfo {author} {\bibfnamefont {P.~M.}\ \bibnamefont {Zimmerman}},\ }\href@noop {} {\bibfield  {journal} {\bibinfo  {journal} {The Journal of Physical Chemistry A}\ }\textbf {\bibinfo {volume} {122}},\ \bibinfo {pages} {2714} (\bibinfo {year} {2018})}\BibitemShut {NoStop}%
\bibitem [{\citenamefont {Campbell}(2019)}]{campbell2019random}%
  \BibitemOpen
  \bibfield  {author} {\bibinfo {author} {\bibfnamefont {E.}~\bibnamefont {Campbell}},\ }\href@noop {} {\bibfield  {journal} {\bibinfo  {journal} {Physical {R}eview {L}etters}\ }\textbf {\bibinfo {volume} {123}},\ \bibinfo {pages} {070503} (\bibinfo {year} {2019})}\BibitemShut {NoStop}%
\bibitem [{\citenamefont {Babbush}\ \emph {et~al.}(2016)\citenamefont {Babbush}, \citenamefont {Berry}, \citenamefont {Kivlichan}, \citenamefont {Wei}, \citenamefont {Love},\ and\ \citenamefont {Aspuru-Guzik}}]{babbush2016exponentially}%
  \BibitemOpen
  \bibfield  {author} {\bibinfo {author} {\bibfnamefont {R.}~\bibnamefont {Babbush}}, \bibinfo {author} {\bibfnamefont {D.~W.}\ \bibnamefont {Berry}}, \bibinfo {author} {\bibfnamefont {I.~D.}\ \bibnamefont {Kivlichan}}, \bibinfo {author} {\bibfnamefont {A.~Y.}\ \bibnamefont {Wei}}, \bibinfo {author} {\bibfnamefont {P.~J.}\ \bibnamefont {Love}}, \ and\ \bibinfo {author} {\bibfnamefont {A.}~\bibnamefont {Aspuru-Guzik}},\ }\href@noop {} {\bibfield  {journal} {\bibinfo  {journal} {New Journal of Physics}\ }\textbf {\bibinfo {volume} {18}},\ \bibinfo {pages} {033032} (\bibinfo {year} {2016})}\BibitemShut {NoStop}%
\bibitem [{\citenamefont {Berry}\ \emph {et~al.}(2019)\citenamefont {Berry}, \citenamefont {Gidney}, \citenamefont {Motta}, \citenamefont {McClean},\ and\ \citenamefont {Babbush}}]{berry2019qubitization}%
  \BibitemOpen
  \bibfield  {author} {\bibinfo {author} {\bibfnamefont {D.~W.}\ \bibnamefont {Berry}}, \bibinfo {author} {\bibfnamefont {C.}~\bibnamefont {Gidney}}, \bibinfo {author} {\bibfnamefont {M.}~\bibnamefont {Motta}}, \bibinfo {author} {\bibfnamefont {J.~R.}\ \bibnamefont {McClean}}, \ and\ \bibinfo {author} {\bibfnamefont {R.}~\bibnamefont {Babbush}},\ }\href@noop {} {\bibfield  {journal} {\bibinfo  {journal} {Quantum}\ }\textbf {\bibinfo {volume} {3}},\ \bibinfo {pages} {208} (\bibinfo {year} {2019})}\BibitemShut {NoStop}%
\bibitem [{\citenamefont {Fowler}\ \emph {et~al.}(2012)\citenamefont {Fowler}, \citenamefont {Mariantoni}, \citenamefont {Martinis},\ and\ \citenamefont {Cleland}}]{fowler2012surface}%
  \BibitemOpen
  \bibfield  {author} {\bibinfo {author} {\bibfnamefont {A.~G.}\ \bibnamefont {Fowler}}, \bibinfo {author} {\bibfnamefont {M.}~\bibnamefont {Mariantoni}}, \bibinfo {author} {\bibfnamefont {J.~M.}\ \bibnamefont {Martinis}}, \ and\ \bibinfo {author} {\bibfnamefont {A.~N.}\ \bibnamefont {Cleland}},\ }\href@noop {} {\bibfield  {journal} {\bibinfo  {journal} {Physical Review A}\ }\textbf {\bibinfo {volume} {86}},\ \bibinfo {pages} {032324} (\bibinfo {year} {2012})}\BibitemShut {NoStop}%
\bibitem [{\citenamefont {Litinski}(2019{\natexlab{a}})}]{litinski2019magic}%
  \BibitemOpen
  \bibfield  {author} {\bibinfo {author} {\bibfnamefont {D.}~\bibnamefont {Litinski}},\ }\href@noop {} {\bibfield  {journal} {\bibinfo  {journal} {Quantum}\ }\textbf {\bibinfo {volume} {3}},\ \bibinfo {pages} {205} (\bibinfo {year} {2019}{\natexlab{a}})}\BibitemShut {NoStop}%
\bibitem [{\citenamefont {Webber}\ \emph {et~al.}(2022)\citenamefont {Webber}, \citenamefont {Elfving}, \citenamefont {Weidt},\ and\ \citenamefont {Hensinger}}]{webber2022impact}%
  \BibitemOpen
  \bibfield  {author} {\bibinfo {author} {\bibfnamefont {M.}~\bibnamefont {Webber}}, \bibinfo {author} {\bibfnamefont {V.}~\bibnamefont {Elfving}}, \bibinfo {author} {\bibfnamefont {S.}~\bibnamefont {Weidt}}, \ and\ \bibinfo {author} {\bibfnamefont {W.~K.}\ \bibnamefont {Hensinger}},\ }\href@noop {} {\bibfield  {journal} {\bibinfo  {journal} {AVS Quantum Science}\ }\textbf {\bibinfo {volume} {4}},\ \bibinfo {pages} {013801} (\bibinfo {year} {2022})}\BibitemShut {NoStop}%
\bibitem [{\citenamefont {Fowler}\ and\ \citenamefont {Gidney}(2018)}]{fowler2018low}%
  \BibitemOpen
  \bibfield  {author} {\bibinfo {author} {\bibfnamefont {A.~G.}\ \bibnamefont {Fowler}}\ and\ \bibinfo {author} {\bibfnamefont {C.}~\bibnamefont {Gidney}},\ }\href@noop {} {\bibfield  {journal} {\bibinfo  {journal} {arXiv preprint arXiv:1808.06709}\ } (\bibinfo {year} {2018})}\BibitemShut {NoStop}%
\bibitem [{\citenamefont {Litinski}(2019{\natexlab{b}})}]{litinski2019game}%
  \BibitemOpen
  \bibfield  {author} {\bibinfo {author} {\bibfnamefont {D.}~\bibnamefont {Litinski}},\ }\href@noop {} {\bibfield  {journal} {\bibinfo  {journal} {Quantum}\ }\textbf {\bibinfo {volume} {3}},\ \bibinfo {pages} {128} (\bibinfo {year} {2019}{\natexlab{b}})}\BibitemShut {NoStop}%
\bibitem [{\citenamefont {Litinski}\ \emph {et~al.}(2017)\citenamefont {Litinski}, \citenamefont {Kesselring}, \citenamefont {Eisert},\ and\ \citenamefont {von Oppen}}]{litinski2017combining}%
  \BibitemOpen
  \bibfield  {author} {\bibinfo {author} {\bibfnamefont {D.}~\bibnamefont {Litinski}}, \bibinfo {author} {\bibfnamefont {M.~S.}\ \bibnamefont {Kesselring}}, \bibinfo {author} {\bibfnamefont {J.}~\bibnamefont {Eisert}}, \ and\ \bibinfo {author} {\bibfnamefont {F.}~\bibnamefont {von Oppen}},\ }\href@noop {} {\bibfield  {journal} {\bibinfo  {journal} {Physical Review X}\ }\textbf {\bibinfo {volume} {7}},\ \bibinfo {pages} {031048} (\bibinfo {year} {2017})}\BibitemShut {NoStop}%
\bibitem [{\citenamefont {Beverland}\ \emph {et~al.}(2016)\citenamefont {Beverland}, \citenamefont {Buerschaper}, \citenamefont {Koenig}, \citenamefont {Pastawski}, \citenamefont {Preskill},\ and\ \citenamefont {Sijher}}]{beverland2016protected}%
  \BibitemOpen
  \bibfield  {author} {\bibinfo {author} {\bibfnamefont {M.~E.}\ \bibnamefont {Beverland}}, \bibinfo {author} {\bibfnamefont {O.}~\bibnamefont {Buerschaper}}, \bibinfo {author} {\bibfnamefont {R.}~\bibnamefont {Koenig}}, \bibinfo {author} {\bibfnamefont {F.}~\bibnamefont {Pastawski}}, \bibinfo {author} {\bibfnamefont {J.}~\bibnamefont {Preskill}}, \ and\ \bibinfo {author} {\bibfnamefont {S.}~\bibnamefont {Sijher}},\ }\href@noop {} {\bibfield  {journal} {\bibinfo  {journal} {Journal of Mathematical Physics}\ }\textbf {\bibinfo {volume} {57}},\ \bibinfo {pages} {022201} (\bibinfo {year} {2016})}\BibitemShut {NoStop}%
\bibitem [{\citenamefont {Bravyi}\ and\ \citenamefont {K{\"o}nig}(2013)}]{bravyi2013classification}%
  \BibitemOpen
  \bibfield  {author} {\bibinfo {author} {\bibfnamefont {S.}~\bibnamefont {Bravyi}}\ and\ \bibinfo {author} {\bibfnamefont {R.}~\bibnamefont {K{\"o}nig}},\ }\href@noop {} {\bibfield  {journal} {\bibinfo  {journal} {Physical {R}eview {L}etters}\ }\textbf {\bibinfo {volume} {110}},\ \bibinfo {pages} {170503} (\bibinfo {year} {2013})}\BibitemShut {NoStop}%
\bibitem [{\citenamefont {Selinger}(2015)}]{selinger2015efficient}%
  \BibitemOpen
  \bibfield  {author} {\bibinfo {author} {\bibfnamefont {P.}~\bibnamefont {Selinger}},\ }\href@noop {} {\bibfield  {journal} {\bibinfo  {journal} {Quantum Info. Comput.}\ }\textbf {\bibinfo {volume} {15}},\ \bibinfo {pages} {159–180} (\bibinfo {year} {2015})}\BibitemShut {NoStop}%
\bibitem [{\citenamefont {Hastings}(2016)}]{hastings2016turning}%
  \BibitemOpen
  \bibfield  {author} {\bibinfo {author} {\bibfnamefont {M.~B.}\ \bibnamefont {Hastings}},\ }\href@noop {} {\bibfield  {journal} {\bibinfo  {journal} {arXiv preprint arXiv:1612.01011}\ } (\bibinfo {year} {2016})}\BibitemShut {NoStop}%
\bibitem [{\citenamefont {Johnson~III}(2022)}]{nist_data_base}%
  \BibitemOpen
  \bibfield  {author} {\bibinfo {author} {\bibfnamefont {R.~D.}\ \bibnamefont {Johnson~III}},\ }\href@noop {} {\  (\bibinfo {year} {2022})},\ \bibinfo {note} {\url{https://cccbdb.nist.gov}}\BibitemShut {NoStop}%
\bibitem [{\citenamefont {Stewart}(1970)}]{stewart1970}%
  \BibitemOpen
  \bibfield  {author} {\bibinfo {author} {\bibfnamefont {R.~F.}\ \bibnamefont {Stewart}},\ }\href@noop {} {\bibfield  {journal} {\bibinfo  {journal} {The Journal of Chemical Physics}\ }\textbf {\bibinfo {volume} {52}},\ \bibinfo {pages} {431} (\bibinfo {year} {1970})}\BibitemShut {NoStop}%
\bibitem [{\citenamefont {Ditchfield}\ \emph {et~al.}(1971)\citenamefont {Ditchfield}, \citenamefont {Hehre},\ and\ \citenamefont {Pople}}]{ditchfield1971}%
  \BibitemOpen
  \bibfield  {author} {\bibinfo {author} {\bibfnamefont {R.}~\bibnamefont {Ditchfield}}, \bibinfo {author} {\bibfnamefont {W.~J.}\ \bibnamefont {Hehre}}, \ and\ \bibinfo {author} {\bibfnamefont {J.~A.}\ \bibnamefont {Pople}},\ }\href@noop {} {\bibfield  {journal} {\bibinfo  {journal} {The Journal of Chemical Physics}\ }\textbf {\bibinfo {volume} {54}},\ \bibinfo {pages} {724} (\bibinfo {year} {1971})}\BibitemShut {NoStop}%
\bibitem [{\citenamefont {Dunning}(1989)}]{dunning1989}%
  \BibitemOpen
  \bibfield  {author} {\bibinfo {author} {\bibfnamefont {T.~H.}\ \bibnamefont {Dunning}},\ }\href@noop {} {\bibfield  {journal} {\bibinfo  {journal} {The Journal of Chemical Physics}\ }\textbf {\bibinfo {volume} {90}},\ \bibinfo {pages} {1007} (\bibinfo {year} {1989})}\BibitemShut {NoStop}%
\bibitem [{\citenamefont {Li}\ \emph {et~al.}(2019)\citenamefont {Li}, \citenamefont {Li}, \citenamefont {Dattani}, \citenamefont {Umrigar},\ and\ \citenamefont {Chan}}]{li2019electronic}%
  \BibitemOpen
  \bibfield  {author} {\bibinfo {author} {\bibfnamefont {Z.}~\bibnamefont {Li}}, \bibinfo {author} {\bibfnamefont {J.}~\bibnamefont {Li}}, \bibinfo {author} {\bibfnamefont {N.~S.}\ \bibnamefont {Dattani}}, \bibinfo {author} {\bibfnamefont {C.}~\bibnamefont {Umrigar}}, \ and\ \bibinfo {author} {\bibfnamefont {G.~K.-L.}\ \bibnamefont {Chan}},\ }\href@noop {} {\bibfield  {journal} {\bibinfo  {journal} {The Journal of Chemical Physics}\ }\textbf {\bibinfo {volume} {150}},\ \bibinfo {pages} {024302} (\bibinfo {year} {2019})}\BibitemShut {NoStop}%
\bibitem [{\citenamefont {Chen}\ \emph {et~al.}(2021)\citenamefont {Chen}, \citenamefont {Satzinger}, \citenamefont {Atalaya}, \citenamefont {Korotkov}, \citenamefont {Dunsworth}, \citenamefont {Sank}, \citenamefont {Quintana}, \citenamefont {McEwen}, \citenamefont {Barends}, \citenamefont {Klimov} \emph {et~al.}}]{chen2021exponential}%
  \BibitemOpen
  \bibfield  {author} {\bibinfo {author} {\bibfnamefont {Z.}~\bibnamefont {Chen}}, \bibinfo {author} {\bibfnamefont {K.~J.}\ \bibnamefont {Satzinger}}, \bibinfo {author} {\bibfnamefont {J.}~\bibnamefont {Atalaya}}, \bibinfo {author} {\bibfnamefont {A.~N.}\ \bibnamefont {Korotkov}}, \bibinfo {author} {\bibfnamefont {A.}~\bibnamefont {Dunsworth}}, \bibinfo {author} {\bibfnamefont {D.}~\bibnamefont {Sank}}, \bibinfo {author} {\bibfnamefont {C.}~\bibnamefont {Quintana}}, \bibinfo {author} {\bibfnamefont {M.}~\bibnamefont {McEwen}}, \bibinfo {author} {\bibfnamefont {R.}~\bibnamefont {Barends}}, \bibinfo {author} {\bibfnamefont {P.~V.}\ \bibnamefont {Klimov}},  \emph {et~al.},\ }\href@noop {} {\bibfield  {journal} {\bibinfo  {journal} {arXiv preprint arXiv:2102.06132}\ } (\bibinfo {year} {2021})}\BibitemShut {NoStop}%
\bibitem [{\citenamefont {Ryan-Anderson}\ \emph {et~al.}(2021)\citenamefont {Ryan-Anderson}, \citenamefont {Bohnet}, \citenamefont {Lee}, \citenamefont {Gresh}, \citenamefont {Hankin}, \citenamefont {Gaebler}, \citenamefont {Francois}, \citenamefont {Chernoguzov}, \citenamefont {Lucchetti}, \citenamefont {Brown} \emph {et~al.}}]{ryan2021realization}%
  \BibitemOpen
  \bibfield  {author} {\bibinfo {author} {\bibfnamefont {C.}~\bibnamefont {Ryan-Anderson}}, \bibinfo {author} {\bibfnamefont {J.~G.}\ \bibnamefont {Bohnet}}, \bibinfo {author} {\bibfnamefont {K.}~\bibnamefont {Lee}}, \bibinfo {author} {\bibfnamefont {D.}~\bibnamefont {Gresh}}, \bibinfo {author} {\bibfnamefont {A.}~\bibnamefont {Hankin}}, \bibinfo {author} {\bibfnamefont {J.}~\bibnamefont {Gaebler}}, \bibinfo {author} {\bibfnamefont {D.}~\bibnamefont {Francois}}, \bibinfo {author} {\bibfnamefont {A.}~\bibnamefont {Chernoguzov}}, \bibinfo {author} {\bibfnamefont {D.}~\bibnamefont {Lucchetti}}, \bibinfo {author} {\bibfnamefont {N.~C.}\ \bibnamefont {Brown}},  \emph {et~al.},\ }\href@noop {} {\bibfield  {journal} {\bibinfo  {journal} {Physical Review X}\ }\textbf {\bibinfo {volume} {11}},\ \bibinfo {pages} {041058} (\bibinfo {year} {2021})}\BibitemShut {NoStop}%
\bibitem [{\citenamefont {Lee}\ \emph {et~al.}(2021)\citenamefont {Lee}, \citenamefont {Berry}, \citenamefont {Gidney}, \citenamefont {Huggins}, \citenamefont {McClean}, \citenamefont {Wiebe},\ and\ \citenamefont {Babbush}}]{lee2021even}%
  \BibitemOpen
  \bibfield  {author} {\bibinfo {author} {\bibfnamefont {J.}~\bibnamefont {Lee}}, \bibinfo {author} {\bibfnamefont {D.~W.}\ \bibnamefont {Berry}}, \bibinfo {author} {\bibfnamefont {C.}~\bibnamefont {Gidney}}, \bibinfo {author} {\bibfnamefont {W.~J.}\ \bibnamefont {Huggins}}, \bibinfo {author} {\bibfnamefont {J.~R.}\ \bibnamefont {McClean}}, \bibinfo {author} {\bibfnamefont {N.}~\bibnamefont {Wiebe}}, \ and\ \bibinfo {author} {\bibfnamefont {R.}~\bibnamefont {Babbush}},\ }\href {\doibase 10.1103/PRXQuantum.2.030305} {\bibfield  {journal} {\bibinfo  {journal} {PRX Quantum}\ }\textbf {\bibinfo {volume} {2}},\ \bibinfo {pages} {030305} (\bibinfo {year} {2021})}\BibitemShut {NoStop}%
\end{thebibliography}%


\section*{Tables}

\begin{table}[!ht]
\caption{The number of T gates for FeMoco molecule as estimated by various sources. The Hamiltonians were taken from the repository of \cite{lee2021even} used in \cite{reiher2017elucidating} and \cite{li2019electronic}. See text for definitions of the algorithms.}\label{t7:FeMoco}
\vspace{2mm}
\centering
\resizebox{8cm}{!}{
\begin{tabular}{|c|c|c|}
\hline
FeMoco active space: & Reiher~\cite{reiher2017elucidating} & Li~\cite{li2019electronic} \\ \hline
This work: & 1.45e15  &  3.73e16  \\ \hline
TF-Trot~\cite{casares2022tfermion}: & 7.34e23 & 3.62e23 \\ \hline
TF-SF~\cite{casares2022tfermion}: & 2.36e13 & 2.17e13 \\ \hline
SF~\cite{berry2019qubitization}   & 4.8e12 & 3.9e12 \\ \hline
\cite{reiher2017elucidating}: & 1.10e15 &  \\ \hline
\end{tabular}}
\end{table}

\begin{table}[!ht]
\caption{Hardware parameters used in the error correction calculations. *Error rates used are two orders of magnitude lower than reported in Refs.~\cite{chen2021exponential} and~\cite{ryan2021realization}, due to the need to be below the standard surface code threshold of 1e-3. $^\dagger$ The cycle time of 70ms for the trapped ion is a third of that reported in Ref.~\cite{ryan2021realization} as their protocol was a more complicated color code protocol.}\label{t7:hw_params}
\vspace{2mm}
\centering
\resizebox{8cm}{!}{
\begin{tabular}{|c|c|c|}
\hline
Hardware & Cycle Time & Error rate \\ \hline
Superconducting Qubit~\cite{chen2021exponential} & 1$\mu$s  &  5e-4*  \\ \hline
Trapped Ion~\cite{ryan2021realization} & 70ms$^\dagger$ & 3e-5* \\ \hline
\end{tabular}}
\end{table}



\section*{Figure captions}


   


\end{document}